\documentclass[a4paper,fleqn]{cas-dc}

\usepackage[numbers,sort&compress]{natbib}

\def\tsc#1{\csdef{#1}{\textsc{\lowercase{#1}}\xspace}}
\tsc{WGM}
\tsc{QE}
\tsc{EP}
\tsc{PMS}
\tsc{BEC}
\tsc{DE}

\begin{document}
\let\WriteBookmarks\relax
\def\floatpagepagefraction{1}
\def\textpagefraction{.001}

\shorttitle{Automated Cervical Cancer Detection}

\shortauthors{Maben et al.}

\title [mode = title]{Automated Cervical Cancer Detection through Visual Inspection with Acetic Acid in Resource-Poor Settings with Lightweight Deep Learning Models Deployed on an Android Device}                      



%
\author[1]{Leander Melroy Maben}[prefix=]


\ead{lmaben@andrew.cmu.edu}


\affiliation[1]{organization={School of Computer Science, Carnegie Mellon University},
    city={Pittsburgh},
    postcode={15232}, 
    state={PA},
    country={USA}}

\author[2]{Keerthana Prasad}[ orcid=0000-0002-4138-6814]
\cormark[1]
\ead{Keerthana.prasad@manipal.edu}

\affiliation[2]{organization={Manipal School of Information Sciences, Manipal Academy of Higher Education},
    city={Manipal},
    postcode={576104}, 
    state={Karnataka},
    country={India}}

\author[3]{Shyamala Guruvare}[prefix=]
\ead{shyamala.g@manipal.edu}

\affiliation[3]{organization={Department of Obstetrics and Gynecology, Kasturba Medical
College},
    city={Manipal},
    postcode={576104}, 
    state={Karnataka},
    country={India}}

\author[2,4]{Vidya Kudva}[prefix=]
\ead{vidyakudva@nitte.edu.in}

\affiliation[4]{organization={NMAM Institute of Technology, Nitte},
    city={Karkala},
    postcode={574110}, 
    state={Karnataka},
    country={India}}

\author[5]{P C Siddalingaswamy}[ orcid=0000-0002-7597-7591]
\ead{pcs.swamy@manipal.edu}

\affiliation[5]{organization={Manipal Institute of Technology, Manipal Academy of Higher Education},
    city={Manipal},
    postcode={576104}, 
    state={Karnataka},
    country={India}}

\cortext[cor1]{Corresponding author}



\begin{abstract}
Cervical cancer is among the most commonly occurring cancer among women and claims a huge number of lives in low and middle-income countries despite being relatively easy to treat. Several studies have shown that public screening programs can bring down cervical cancer incidence and mortality rates significantly. While several screening tests are available, visual inspection with acetic acid (VIA)  presents itself as the most viable option for low-resource settings due to the affordability and simplicity of performing the test. VIA requires a trained medical professional to interpret the test and is subjective in nature. Automating VIA using AI eliminates subjectivity and would allow shifting of the task to less trained health workers. Task shifting with AI would help further expedite screening programs in low-resource settings. In our work, we propose a lightweight deep learning algorithm that includes EfficientDet-Lite3 as the Region of Interest (ROI) detector and a MobileNet-V2 based model for classification. These models would be deployed on an android-based device that can operate remotely and provide almost instant results without the requirement of highly-trained medical professionals, labs, sophisticated infrastructure, or internet connectivity. The classification model gives an accuracy of 92.31\%, a sensitivity of 98.24\%, and a specificity of 88.37\% on the test dataset and presents itself as a promising automated low-resource screening approach. 

\end{abstract}




\maketitle

\section{Introduction}

Cervical cancer is the fourth most common cancer among women globally. It caused an estimated 342,000 deaths in 2020 \cite{who1} and represents 6.5\% of all female cancers\cite{togetherforhealthGlobalBurden}. The large number of deaths is of major concern, especially because cervical cancer is one of the most successfully treatable cancers when it is detected at an early stage and managed effectively.

Cervical cancer develops in a woman’s cervix, which is the entrance to the uterus from the vagina \cite{whoCervicalCancer}. Almost all cervical cancers are linked to infection with high-risk human papillomaviruses (HPV), an extremely common virus that transmits through sexual contact \cite{togetherforhealthGlobalBurden}.
Persistent infection with high-risk HPV types may lead to the development of precursor lesions of the cervix, called CIN (cervical intraepithelial neoplasia), which represent epithelial cellular changes. CIN can be graded as CIN-1 (mild), CIN-2 (moderate), and CIN-3 (severe). Moderate and severe CINs are more likely to develop into invasive cancer, although a proportion of even these grades of CIN may regress or persist. Invasive cancer develops from CIN progressively from mild to moderate to severe and then cancer. This progression from pre-cancer to cancer takes 7 to 20 years \cite{mishra2011yx, Balasubramaniam2019yr}. The long lag between pre-cancer and cancer can be effectively used to screen, detect and treat pre-cancer before progression to cancer \cite{nihBackground}.

About 90\% of new cervical cancer cases and deaths in 2020 occurred in low and medium-income countries \cite{who1}. Moreover, cervical cancer was found to be the most common type of cancer among women in 36 countries of the same income groups. Unsurprisingly, the low-resource countries which bore 90\% of cervical cancer deaths had limited prevention, screening, and treatment services \cite{togetherforhealthGlobalBurden}. These staggering statistics highlight the need for a cost and resource-effective way to screen women for cervical cancer in highly rural and remote settings.

\begin{figure}
	\centering
		\includegraphics[scale=.7]{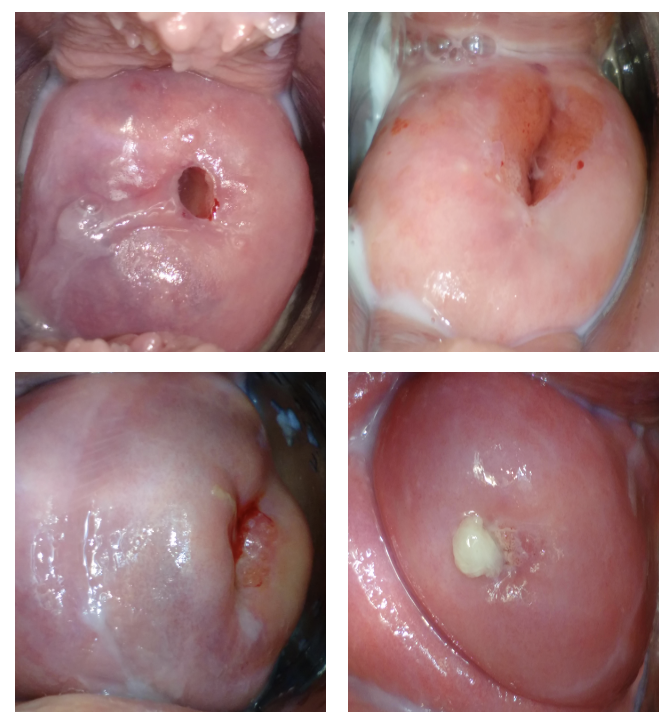}
	\caption{Sample VIA-negative images. The figure shows cervixes that do not have aceto-white regions.}
	\label{FIG:via-neg}
\end{figure}

\begin{figure}
	\centering
		\includegraphics[scale=.7]{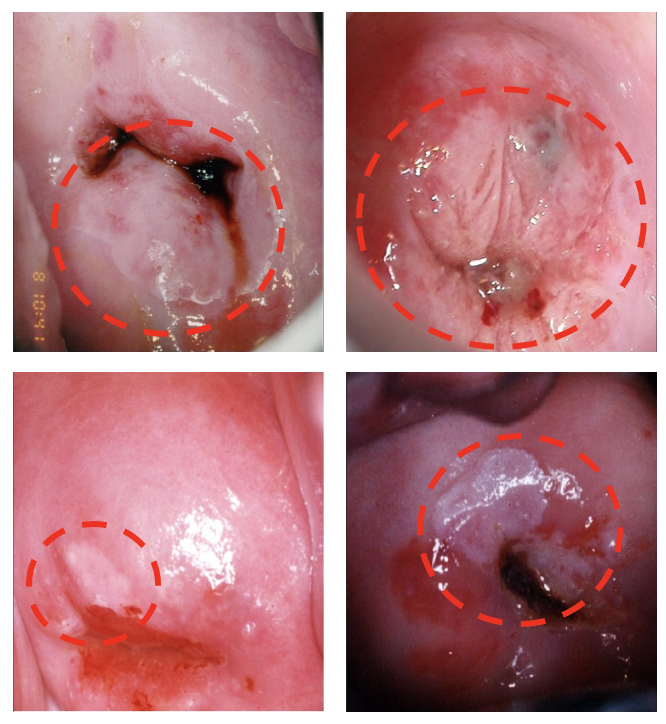}
	\caption{Sample VIA-positive images \cite{iarcatlas}. The figure shows cervixes that have aceto-white regions. These regions have been marked with a dashed red circle.}
	\label{FIG:via-pos}
\end{figure}

Conventionally used screening practices for cervical cancer include cytology-based tests like Papanicolaou (Pap) smear or the newer liquid-based cytology technique, HPV tests, and visual inspection \cite{BEDELL202028}. Soon after its discovery, the pap smear became the gold standard for cervical cancer screening and is still used in primary screening. Cells from the transformation zone are collected and transferred to a liquid preservative, followed by automated processing or inspection under the microscope for liquid-based cytology and conventional pap smear, respectively \cite{BEDELL202028, nihPapanicolaouSmear}. Pap smear and liquid-based cytology do not show considerable differences in sensitivity and specificity in detecting CIN. Pap smear has demonstrated consistent specificity of 98\% but lower and variable sensitivity of 55\%-80\%. It relies on repeated screening throughout a woman’s lifetime to balance the low sensitivity. Cytology-based screening is not feasible for low-resource settings due to the requirement of electricity for microscopes, supplies for testing, trained cytopathologists, and repeated screening. This is coupled with the difficulty of patient follow-up in case of positive results and the poor sensitivity and specificity for high-grade lesions recorded in developing countries \cite{BEDELL202028}. HPV tests can be used to detect HPVs. HPV tests also come with a self-swab variant where a patient can collect the vaginal sample by herself without the need for a medical professional. Testing for high-risk HPVs has been shown to achieve higher sensitivity than cytology-based tests in detecting high-grade CIN and can thus help prevent cervical cancer more effectively \cite{Thomsen2020-er}.  HPV testing is unsuitable for low-resource settings due to the high cost and the requirement of a laboratory. Visual Inspection with Acetic acid (VIA) is one of the techniques to screen for cervical cancer based on visual inspection. It involves visual inspection of the cervix after applying 3\% to 5\% acetic acid. Application of acetic acid at 3\% to 5\% causes reversible coagulation of cellular protein. Maximal coagulation occurs in areas with dysplasia or invasive cancer due to higher concentrations of proteins, making these areas appear acetowhite \cite{mishra2011yx}. Examples of VIA-negative and VIA-positive cervix images are shown in figure \ref{FIG:via-neg} and figure \ref{FIG:via-pos}, respectively. VIA is low-cost, safe, and can be performed by a wide range of medical providers owing to its simplicity, making it suitable for use in low-resource settings. Results are provided instantly without the need for laboratory processing. Reviews have reported a sensitivity of 84\% and specificity of 82\% in detecting high-grade dysplasia \cite{BEDELL202028}. 

While VIA is a promising technique to curb cervical cancer mortality rates in remote areas, it still requires a trained professional to interpret the results. This poses a problem as the lack of professionals is often a major issue in resource-poor areas. Moreover, the accuracy of the test could vary based on the professional’s skills, as the test is subjective. Computer-based processing can go a long way in helping shift the task from highly trained professionals to less trained health workers and in neutralizing the variability in test performance as a result of variations in human skill. The aim of this study is to develop an algorithm that can process cervix images after the application of 3\% to 5\% acetic acid and detect the presence of pre-cancerous or cancerous lesions. Such an algorithm deployed on a portable android device can be used by health workers- even workers without intensive training to interpret VIA results - in remote areas to screen for cervical cancer.

\section{Related Work}

Cervical cancer incidence and mortality rates can be drastically reduced through well-organized screening programs. In a study on cervical cancer incidence and mortality rates in Canada based on national data (1932-2006), Dickinson et al. \cite{Dickinson2012}  point out that from a high mortality rate in the 1950s, the mortality rate in Canada has reached one of the lowest in the world (the risk of dying from cervical cancer is 0.22\%). While they attribute the early decrease from 1950 to 1970 to improved treatment methods, the recent decrease in incidence and mortality is credited to high screening rates. Robles et al. \cite{Robles1996-wj} impute the absence of a decline in cervical cancer mortality rates in Latin America as opposed to trends in countries like Canada and the United States to the lack of screening and screening shortcomings in these countries. Mandelblatt et al. \cite{mandel} use a population-based simulation to evaluate the costs and benefits of using seven screening techniques, including VIA, HPV testing, Pap Smear, and their combinations, in a developing country, namely Thailand. They find that any of the seven strategies would be better than having no well-organized screening program and reduce mortality by up to 58\%.

Several tests for cervical cancer screening are available, like cytology-based tests, HPV tests, and visual inspection. A cross-sectional study by Nkwabong et al. \cite{Nkwabong2019-kt} shows that pap smear has a low sensitivity of 55.5\% and a specificity of 75\%. Liquid-based cytology, a relatively newer type of cytology test compared to the pap smear, has been shown to be superior to conventional pap smear only in terms of fewer unsatisfactory smears in a study by Pankaj et al. \cite{Pankaj2018-vz}. The same study concludes that conventional pap is more suitable in a low-resource setting considering the economic implications of liquid-based cytology. Using HPV screening as a primary test has the benefit of higher sensitivity (95\%), better reassurance from a negative test, and safe prolongation of screening intervals \cite{Bhatla2020-gz}. However, according to Ghisu \cite{Ghisu2021-lo}, the superiority of HPV testing over Pap smear is proven only under defined study conditions after two or three screening rounds. If screening is conducted opportunistically, there is a risk of lost follow-up. Furthermore, pap has a higher specificity than HPV testing, and the sensitivity can be improved to match HPV tests by using additional immunohiostochemical tests. Pap is still considered very effective in countries with sufficient resources. Sauvaget et al. \cite{SAUVAGET201114} report VIA to have a sensitivity of 80\% and a specificity of 92\% and conclude that VIA is an efficient, cost-effective alternative to cytological testing in low-resource areas. In a study by Bobdey et al. \cite{Bobdey2016-vu}, they conclude that due to the lack of resources in countries like India, large-scale implementation of high-quality cytological tests may be infeasible and that visual inspection with acetic acid(VIA) or Lugol’s iodine(VILI) should be adopted in primary health-care. Using computer-based models to evaluate the cost-effectiveness of various screening strategies in five developing countries, Goldie et al. \cite{Goldie2005-ci} find that screening once in a woman’s lifetime at the age of 35 with VIA or HPV test reduced the lifetime risk of cancer by 25 to 36 percent and would cost less than \$500 per year of life saved. In the study involving seven screening techniques, including VIA, HPV testing, pap smear, and their combinations, Mandelblatt et al. \cite{mandel} discovered that performing VIA in five-year intervals followed by immediate treatment in case of abnormalities for women aged 35-55 would save the most lives and be the least expensive approach.

A major concern with VIA is that it is subjective and is prone to high inter-observer variability. One way to mitigate this drawback is to use automated systems and algorithms to perform the test. These systems would make evaluations using a standardized approach and eliminate subjectivity. A good deal of research has been done on automating the process of visual inspection.

 The quality of the cervix images plays a crucial role in proper diagnosis. To that end, research has been done to detect cervix image quality automatically. Guo et al. \cite{diagnostics10070451} propose an ensemble of three deep learning frameworks, including RetinaNet, Deep SVDD, and a customized CNN, to detect whether the smartphone-captured images adequately contain the cervix. They evaluate the performance of the individual models as well as the ensemble. The concern of image focus was addressed in a paper by Guo et al. \cite{8834495}, presenting three deep learning models to evaluate the sharpness of cervix images captured using a smartphone.

Depending on the size and location of the squamo-columnar junction (SCJ), the transformation zone (TZ) or cervix is classified into three types \cite{iarcColposcopyDigital}. The treatment plan for the patient would vary depending on the cervix type. Gorantla et al. \cite{8941985} put forward the CervixNet methodology, where they perform cervix type classification using the hierarchical Convolutional Mixture of Experts algorithm after image enhancement and ROI extraction. Another approach to cervix type classification was presented by Aina et al. \cite{9043206} using the lightweight convolutional model SqueezeNet, which is suitable for mobile device deployment.

 There have also been works focused on preprocessing cervix images to make them more suitable for the downstream task of pre-cancer and cancer detection. In the methodology used by Das et al. \cite{6004107}, specular reflection is removed by interpolation, and the ROI is segmented using the k-means clustering algorithm. Kudva et al. \cite{KUDVA2017281} employ a standard deviation filter for specular reflection detection and curvilinear structure enhancement for cervix region (ROI) detection. Liu et al. \cite{LIU201847} use cervix images from before and after acetic acid application to derive a ratio image which is used to facilitate the segmentation of acetowhite regions using the level-set algorithm.

Many research works have aimed at extracting relevant features from the images and then using these features for classification. Several papers have used features related to cervical lesion margins \cite{10.1117/12.651119,1044872}, texture with or without color \cite{inbook,896790,6884827}, and acetowhite features \cite{Li2009-vq}. In the paper by Park et al. \cite{10.1117/1.2830654}, they cluster similar optical patterns and then classify whether the regions obtained contain neoplastic tissue. Li et al. \cite{10.1117/12.708710} propose a multistep process for acetowhite region detection involving color calibration, analysis of cervix anatomy, identification of acetowhite subregions in the squamous region, and color scoring of these subregions to indicate the level of acetowhite.
\begin{figure*}[t]
	\centering
		\includegraphics[scale=.4]{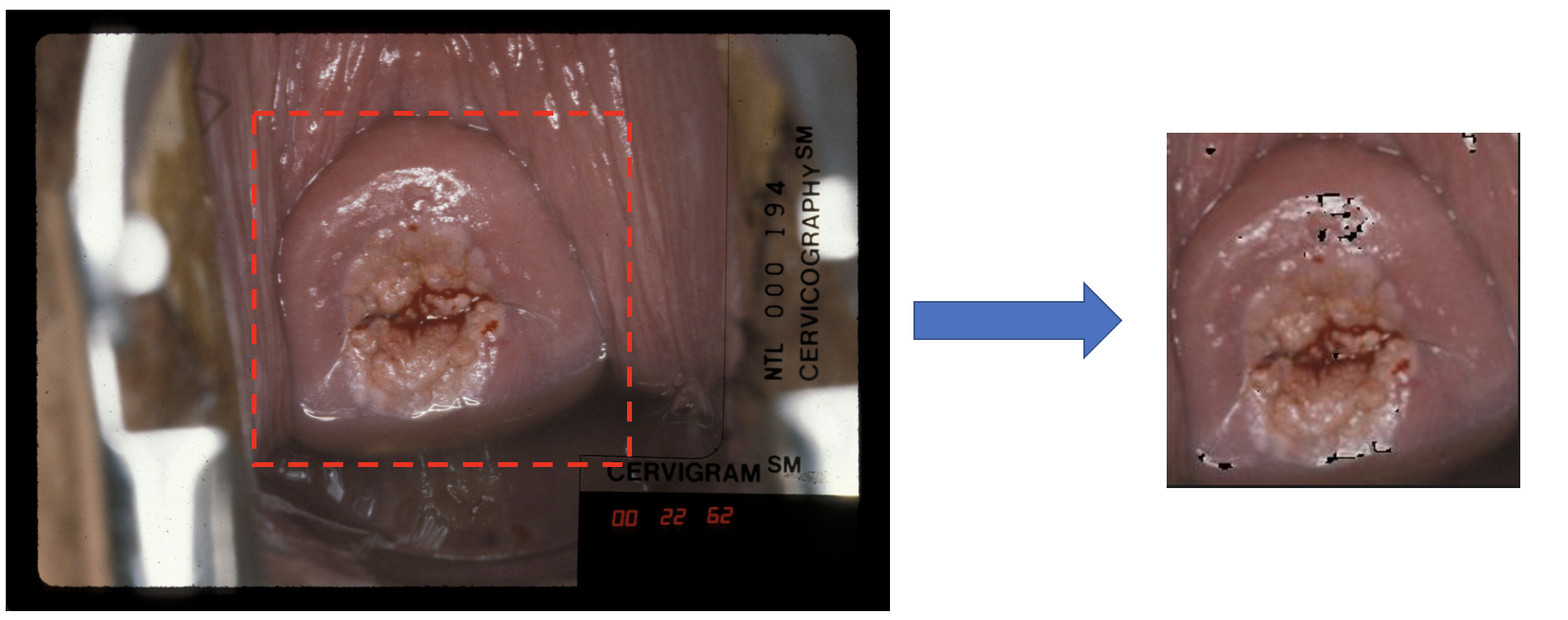}
	\caption{Extracting the region of Interest (ROI) from the cervigram. This process helps isolate the area where acetowhite regions can be formed and eliminates the areas irrelevant for cervical cancer detection.
}
	\label{FIG:roi}
\end{figure*}

With the advent of deep learning, features no longer had to be hand-engineered as the deep learning  models were capable of learning to extract relevant features during training. The deep learning approach has also been prevalent in solving the problem of automating cervical cancer diagnosis. Researchers have proposed using efficient deep learning networks like Colponet to classify colposcope images \cite{Saini2020}, using a shallow CNN to classify manually extracted patches from cervigrams \cite{Kudva2018-yr}, and using techniques like deep metric learning with various models to extract linearly separable image features, which can then be classified using the kNN algorithm \cite{9388707}. In a paper by Kudva et al. \cite{Kudva2020-xe}, a transfer learning technique is presented where filters from networks pre-trained on non-cervix data are manually selected based on inspection of which filters capture features from the cervigrams that are most relevant to cervical cancer detection. These pre-trained filters are then transferred to a new network and used for the classification task. Elakkiya et al. \cite{9473025} detect the cervical spot using a Faster Region-Based CNN  and classify the three types of lesions.The work by Pal et al. \cite{DBLP:journals/corr/abs-2201-07013} demonstrates a way to take advantage of unlabeled data for pre-training as well as comply with data-sharing restrictions through federated self-supervised learning.

Screening based on portable devices like smartphones or other android devices would help significantly improve screening coverage in low-resource areas. Kudva et al. \cite{Kudva2018-sc}, Mansoor et al. \cite{8333312}, and Bae et al. \cite{info:doi/10.2196/16467} propose feature extraction  followed by classification of the cervix images obtained using smartphone/android-based devices after the application of acetic acid. However, these techniques do not harness deep learning capability to automatically learn the best feature for classification. Xue et al. \cite{https://doi.org/10.1002/ijc.33029} demonstrate the usability of the Automated Visual Evaluation (AVE) algorithm on smartphone images. The AVE algorithm uses a Faster-RCNN deep learning model to detect the cervix, extract features and classify the images. Hu et al. \cite{9175863} propose refactoring AVE to use a new deep learning framework to allow a running time of ~30 seconds on a low-end smartphone, and they also present an algorithm to localize the cervix and evaluate image quality. Viñals et al. \cite{diagnostics11040716} use a Samsung Galaxy S5 smartphone to record a video of the cervix after the application of acetic acid. A shallow Artificial Neural Network is employed to generate a probability map which is then used to detect lesion contours. The size of the lesion is used to classify whether they are neoplastic or not. Alyafeai et al. \cite{ALYAFEAI2020112951} use a ResNet-like model for cervix detection and CNN-based models for classification. The lightweight nature of the architecture and speed of the model make it suitable for deployment on mobile devices. While there have been attempts to use smartphones to capture cervix images or videos followed by automated evaluation, work on dedicated portable devices for automated VIA testing with instant result generation is limited. In our work, we propose a dedicated portable device for automated VIA that targets large-scale operation in resource-poor settings equipped with lightweight, locally-run cervix detection and abnormality prediction models.

\section{Methodology}
\subsection{Data}
For the purpose of the study, we collect cervix images from multiple sources. These include 163 images from the Kasturba Medical College (KMC), Manipal, India, 1150 images from the National Institute of Health (NIH), and 177 images from the International Agency for Research on Cancer (IARC). In order to increase the number of positive samples and bring it to a similar number as the negative samples, we augment the positive images by taking a transpose of the images. We add the augmented images to the dataset. Finally, we split the images into train, validation, and test datasets consisting of 1046, 301, and 143 images, respectively. Since multiple images of a given patient are present in the dataset, we take steps to ensure that all images of a given patient fall within the same split. We do this to ensure that images from a given patient do not end up in both the train split and the validation or test split, as this may falsely give a better performance during validation or testing, respectively.

\subsection{ROI Extraction}

Extracting the region of interest (ROI) helps the classification model by presenting only the relevant region - the cervix - as input and eliminating the regions where there is no possibility of occurrence of cancerous lesions. This alleviates the need for the classification model to learn that particular regions do not contribute to the classification process. This simplification becomes even more important given the small amount of data available for classification model training in our study. Figure \ref{FIG:roi} demonstrates ROI extraction on a sample image. 

We train an object detection model to localize the cervix, which is then extracted from the image and used for the classification task. We trained two models for object detection, namely RetinaNet \cite{https://doi.org/10.48550/arxiv.1708.02002} and EfficientDet-Lite3 \cite{https://doi.org/10.48550/arxiv.1911.09070,tensorflow}. We train the models using  cervigram images with annotations for the cervix. The training set consisted of 300 images.

While both RetinaNet and EfficientDet-Lite3 give similar results, we use EfficientDet-Lite3 in the deployed pipeline as its significantly smaller size makes it ideal for use in the proposed device. 
\begin{figure*}
	\centering
		\includegraphics[scale=.5]{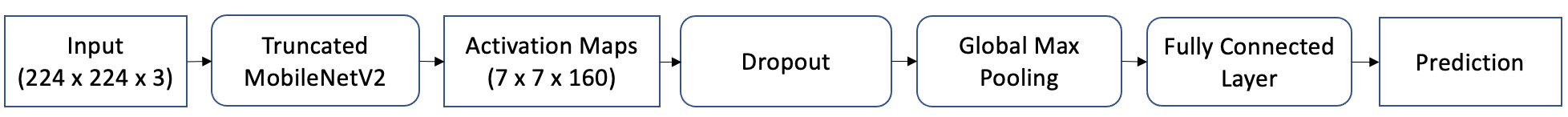}
	\caption{Overview of the image classification model.}
	\label{FIG:trunc_mob}
\end{figure*}

\subsection{Image Classification}
\subsubsection{Image Preprocessing and Augmentation}
We first normalize the images, which is performed as shown in equation \ref{eq:norm}. The mean and standard deviation are computed for the entire training dataset on a per-channel basis, and the equation \ref{eq:norm} is applied to all the pixels in each channel. 
\begin{equation}
     normalized\:image = \frac{image-mean}{standard\:deviation}
    \label{eq:norm}
\end{equation}

Lack of data is a common hindrance for medical-related machine learning tasks. In order to aid in better generalization, we use image augmentation. We augment images dynamically in a random manner during the training process rather than generate an augmented static dataset prior to training. The former method would expose the model to a greater variety of variations than the latter, as the training samples in each epoch would be different due to the random nature of augmentation. We selected a list of augmentations along with accompanying parameters and set the probabilities of application for each of these augmentations. During training, each augmentation is sequentially applied in accordance with the set probabilities. The augmentation details are described in Table \ref{aug}.

\begin{table*}[t]
\caption{Description of Image Augmentation.}\label{aug}
\def\arraystretch{1.5}

\begin{tabular*}{\tblwidth}{@{} L@{\hskip 0.7in}L@{\hskip 0.7in}L@{} }
\toprule
Augmentation & Probability of Application & Description \\
\midrule
Flip & 0.5 & Flips the image horizontally, vertically, or both.\\
Random Contrast & 0.1 & Randomly changes the contrast of the input image.\\
Shift, Scale and Rotate & 0.7 & Randomly shifts, scales, and rotates with a \\
 & &maximum limit of 0.1, 0.1, and $ 180^o $, respectively.\\
Blur & 0.1 & Blurs the image.\\
Grid Shuffle & 0.1 & Splits image into a 3x3 grid and then randomly\\ 
 & &shuffles the cells of this grid.\\
Coarse Dropout & 0.7 & Randomly drops a maximum of 20 rectangular\\ 
& & patches of size 4x4 pixels from the image.\\

\bottomrule
\end{tabular*}
\end{table*}

\begin{table*}[b]
\caption{Architecture of Truncated MobileNetV2. con2D stands for 2-Dimensional Convolution operation.}\label{mobileNet}
\begin{tabular*}{\tblwidth}{@{} CCCCCC@{} }
\toprule
Input Dimensions & Operator & Expansion Factor & Output Channels & Number of Sequential Repetitions & Stride\\
\midrule
$224^2$ x 3 & conv2D & -& 32& 1& 2\\
$112^2$ x 32& bottleneck& 1& 16& 1& 1\\
$112^2$ x 16& bottleneck& 6& 24& 1& 2\\
$56^2$ x 24& bottleneck& 6& 24& 1& 1\\
$56^2$ x 24& bottleneck& 6& 32& 1& 2\\
$28^2$ x 32& bottleneck& 6& 32& 2& 1\\
$28^2$ x 32& bottleneck& 6& 64& 1& 2\\
$14^2$ x 64& bottleneck& 6& 64& 3& 1\\
$14^2$ x 64& bottleneck& 6& 96& 3& 1\\
$14^2$ x 96& bottleneck& 6& 160& 1& 2\\
\bottomrule
\end{tabular*}
\end{table*}

\subsubsection{Model Architecture}
\begin{figure}
	\centering
		\includegraphics[scale=.5]{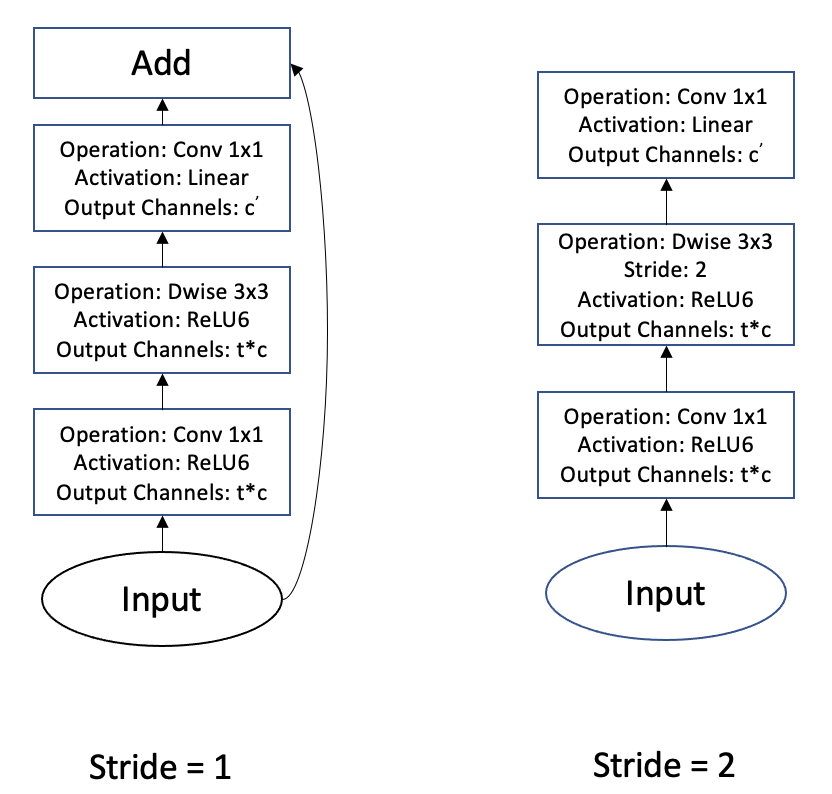}
	\caption{General structure of MobileNetV2 bottlenecks of strides 1 and 2. c represents the number of input channels, c’ represents the number of output channels, and t represents the expansion factor. All convolutions (Conv) and depthwise separable convolutions (Dwise) use a stride of 1 unless specified.
 }
	\label{FIG:bottle}
\end{figure}

For the purpose of image classification, we use a truncated version of MobileNetV2 \cite{https://doi.org/10.48550/arxiv.1801.04381}. An overview of the classification model is depicted in figure \ref{FIG:trunc_mob}. The MobileNetV2 architecture is specifically designed to be suitable for mobile devices in terms of memory and computational costs. It uses depthwise separable convolutions - which are up to 9 times computationally cheaper than traditional convolutions - and Inverse residual structures to make the architecture more efficient. The basic building blocks of MobileNetV2 are bottlenecks, and the overall structure of these bottlenecks is described in figure \ref{FIG:bottle}.

Shallower networks are better at generalizing to new data when the amount of training data is small. The general intuition is that a model with a smaller number of parameters is less likely to overfit the training data than a model with a larger number of parameters. Following this intuition, we exclude the last few layers of MobileNetV2 to reduce the depth of the network. The overall architecture of truncated MobileNetV2 is described in Table \ref{mobileNet}. The outputs from this model are passed through a dropout layer, a global max-pooling layer, and a fully connected layer.

\subsubsection{Training Strategy}

We begin training by loading the pre-trained weights (trained on the ImageNet \cite{5206848} dataset) for the truncated MobileNetV2  and randomly initializing the final fully-connected layer weights following the glorot-uniform distribution. 
We use binary focal loss \cite{https://doi.org/10.48550/arxiv.1708.02002} with the parameter gamma set to 1 as the objective function to be minimized. Moreover, we give a higher weight to the cases with cancer (positive images) to encourage the model to have a higher sensitivity. The strategy we use to stop the model training is early stopping. The validation loss is monitored, and the training is stopped if the validation loss does not decrease for more than 15 epochs. We also set the maximum number of epochs to 70. The model training stops at this limit if the early stopping strategy does not stop training in a lesser number of epochs.
We used the Adam \cite{https://doi.org/10.48550/arxiv.1412.6980} optimizer to optimize the model parameters using an initial learning rate of $10^{-4}$. The learning rate is then repeatedly decreased by a factor of 0.7 during training if the training loss does not decrease for more than 5 epochs.
A dropout layer with a drop rate of 0.2 was used at the end of the truncated MobileNetV2 network in order to add stochasticity and help in better generalization.

\subsection{Proposed Device}

The proposed device, named “Sakhi-Manipal,” is designed as a smartphone capable of capturing cervix images of the required quality. Furthermore, it can store patient records that can then be used to obtain a second opinion. Many features like wifi, sim and SD card, earphone jack, speakers, microphone, volume button, proximity sensor, google assistant, front camera, and fingerprint sensor are disabled for data security reasons and to ensure that the device is not used for unintended purposes. 

It operates under two modes, novice and expert. In the novice mode, it performs inference on the captured cervix image and displays the result. Hence, it acts as an automated screening tool. In the expert mode, it first prompts the expert to input their diagnosis of the image and then provides the result from the detection algorithm. Hence, it acts as a decision support system.

It is highly portable and operates without the requirement of internet connectivity. It is 159.43mm in length, 76mm in width, 9.3mm in height and weighs 198g.

\section{Results}

\begin{figure}
	\centering
		\includegraphics[scale=.5]{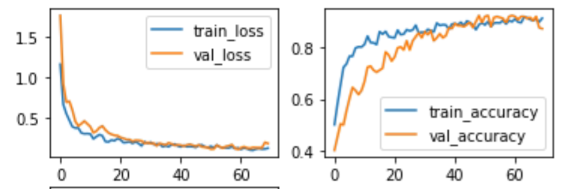}
	\caption{Training curve for the classification model. The plot on the left shows training and validation loss variation over epochs. The plot on the right shows training and validation accuracy variation over epochs. 
 }
	\label{FIG:res}
\end{figure}

The proposed method gives an average accuracy, sensitivity, and specificity of 92.86\%, 93.48\%, and 92.43\%, respectively, for a validation set of 301 images over ten runs. 
We pick the best model in terms of validation metrics from these runs and perform inference on our test dataset of 143 images. We get an accuracy, sensitivity, and specificity of 92.31\%, 98.24\%, and 88.37\%, respectively, for the test set.
The training curves for the training of the best model among the 10 runs are shown in figure \ref{FIG:res}. Training loss is denoted by train\_loss, validation loss is denoted by val\_loss, training accuracy is denoted by train\_accuracy and validation accuracy is denoted by val\_accuracy.

\begin{table*}
\caption{Comparing metrics of different methods with the proposed method.}\label{res_comp}
\begin{tabular*}{\tblwidth}{@{} LLLLL@{} }
\toprule
Authors / Method & Accuracy(\%) & Sensitivity(\%) & Specificity(\%) & Area Under the Curve (AUC)\\
\midrule
Kudva et. al. \cite{Kudva2018-sc} & 97.94 & 99.05 & 97.16& - \\
Automated Visual Evaluation (AVE) \cite{https://doi.org/10.1002/ijc.33029} &  over 90 & 97.7& 85& - \\
Viñals et. al. \cite{diagnostics11040716} & 89 & 90 & 87 & -  \\
Alyafeai et. al. \cite{ALYAFEAI2020112951}&-&-&-&0.68 \\
Bae et. al.\cite{info:doi/10.2196/16467} & 78.3& 75& 80.3& 0.807 \\
Monsur et. al.\cite{8333312} &-&83&79&-\\
Traditional Pap Smear&-&55.5&75&-\\
VIA&-&80&92&-\\
HPV Tests&-&95&-&-\\
Proposed Method& 92.31& 98.24& 88.37& -\\
\bottomrule
\end{tabular*}
\end{table*}

\begin{figure}
	\centering
		\includegraphics[scale=.5]{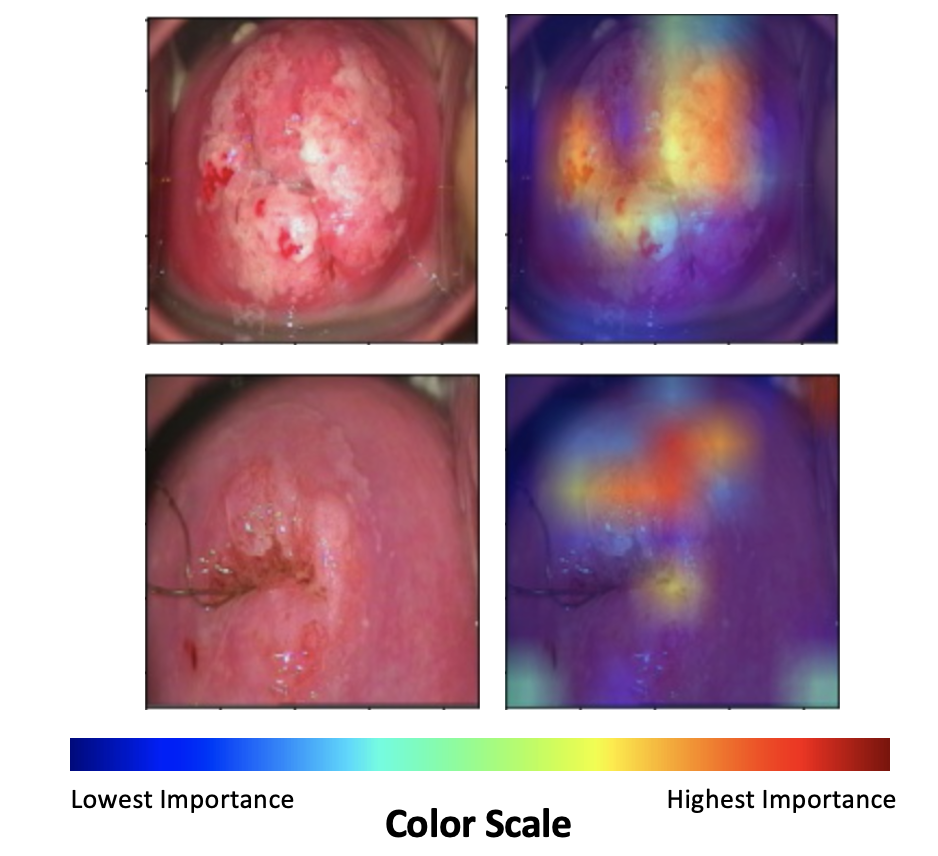}
	\caption{Sample GradCAMs. The parts that play a crucial role in predicting that the cervix is abnormal are highlighted with the more important colors on the scale shown. The less important areas are marked with corresponding colors of low importance.
}
	\label{FIG:gradCam}
\end{figure}

In order to make the classification model more explainable, we use GradCAMs \cite{Selvaraju_2019}. GradCAMs depict the degree to which parts of the input image were used to make an inference by the model. figure \ref{FIG:gradCam} shows GradCAMs for two sample images.

\section{Discussion}

The proposed method attains an accuracy, sensitivity, and specificity of  92.31\%, 98.24\%, and 88.37 \%, respectively, for a test set of 143 images. Our method outperforms the traditional tests of pap smear, VIA, and HPV in terms of sensitivity. This result is promising as high sensitivity is one of the most important characteristics of a screening test. It also has a relatively high specificity compared to the traditional tests, except for traditional VIA, which has a reported specificity of 92\%. Additionally, the trained models are lightweight and are thus suitable for deployment on smaller, portable devices. The ROI extractor and classifier models are 4341 KB and 722 KB in size, respectively. However, one limitation is the relatively small size of the test dataset, which may not cover enough diversity of cervix images that would be encountered in the field. The results, however, can be treated as a reasonable estimate of the algorithm's performance in the real world. Table \ref{res_comp} summarizes the metrics of the proposed method alongside other prominent methods.

Moreover, the proposed device (Sakhi-Manipal) stores the cervix images of patients. It thus acts as a repository of patient test records that would otherwise be lost during a conventional visual examination. This utility enables eliciting expert second opinions if required later, which would, to some extent, alleviate the concern of the unavailability of medical experts in remote areas. In addition to this, it helps shift the tasks of performing VIA from highly trained professionals capable of analyzing the cervix after acetic acid application to less trained health workers who only need to be trained in capturing a good quality cervix image using the device. The burden of analyzing the image is transferred to the AI application in the device.

While Pap smear has been the prevalent technique for screening, and HPV testing is gaining acceptance as a screening method, they have some serious shortcomings in a low-resource setting. Pap smear requires professionals for sample collection, sample processing, and interpretation of the collected samples. In addition, it suffers from inter-observer variability and requires a lab. HPV tests are expensive and require a lab and sophisticated infrastructure. Both Pap Smear and HPV testing take time before the results become available. This delay in diagnosis increases the risk of loss to follow-up.  Performing automated VIA using the device is cost-effective, non-subjective, does not require a lab or expensive infrastructure, gives almost immediate results, and can be performed by health workers with low-level training.

Conventional VIA, pap smear, and HPV testing require pelvic examination, which can be a source of hesitation in women. Our proposed work is no exception to this requirement. On the other hand,  the HPV test using the self-swab method allows the patient to collect the sample without the need for a health worker or a pelvic examination. The high sensitivity of the HPV test coupled with the convenience of self-swab makes this test the ideal way to screen for cervical cancer, given that the necessary resources are available. The drawback, however, is that these tests are very expensive and are not feasible for screening programs in low and low-middle-income countries (LMIC). In addition, the reduction of costs of the HPV test due to technological advancements or other efforts is unlikely to occur at a pace suitable to meet World Health Organization's (WHO) cervical cancer elimination policy. The HPV self-swab test can, however, be offered as an alternative option in the screening programs as women from higher socio-economic strata might find it more acceptable and be able to afford the cost. It should also be noted that some groups prefer the test to be conducted by a medical worker, and hence the impact of hesitancy towards pelvic examination on the overall rate of screening is a subject that needs further investigation.

In medical applications, the security of patient data is an essential consideration. The proposed device has wifi disabled to mitigate security concerns that arise due to it. The transfer of data to another storage unit is through a wired connection. We also take steps to eliminate dependence on internet connectivity. To that end, the device performs computations and predictions locally rather than through a cloud-based platform.

Finally, in order to implement feasible public health systems in LMICs, the screening approaches need to be economical, have less dependence on professionals with a high level of training, and be implementable in remote settings. It is also beneficial for the approach to give results within short time durations to prevent loss to follow-up and enable administering immediate treatment to the patient if the facility is available on-site. AI eliminates expert participation in the cost-effective, quick, and easily implementable VIA; thus, AI-aided VIA presents itself as the best choice for cervical cancer screening in Public Health Programs despite HPV using self-swab being a more convenient and sensitive test.  Further, the WHO initiative to eliminate cervical cancer aims to screen at least 70\% women twice during their lifetime using a high-performance test \cite{world_health_organization}. The presented work would be conducive to this volume of screening in the short time frame in low-resource countries, while HPV testing would be infeasible due to the high costs involved and the infrastructure required.

\section{Conclusion}

In our work, we propose a lightweight automated cervical cancer detection model deployed on a custom android-based device. The proposed method achieves an accuracy of 92.31\% with a sensitivity of 98.24\% and a specificity of 88.37\% on a test dataset of 3 images. Based on results obtained on the specified test dataset, it outperforms the pap smear, conventional VIA, and HPV tests as well as the AVE technique in terms of both sensitivity and specificity. It should be noted that the HPV test is perhaps the ideal screening method due to its high sensitivity and the elimination of hesitancy in the self-swab variant of the test. However, it is the requirement of sophisticated infrastructure and the high cost of the test that makes it unsuitable for public screening in low-resource settings. Furthermore, it is unlikely that the cost can be reduced at the required pace to meet the WHO's cervical cancer elimination goals. That being said, in high-income countries, the HPV test would probably be the best option for primary screening. The long-established pap smear test does not present itself as the ideal test due to infrastructure  requirements as well as poor sensitivity. The affordability, low-resource requirements, and significant sensitivity present VIA as the best solution for primary screening. VIA assisted by AI completely eliminates the subjectivity involved with the test and shifts the tasks from highly trained professionals to less-trained workers. With the capability of producing almost instant results, no requirement of internet connection or skilled workers, and the affordability of screening with the proposed device, Sakhi-Manipal shows promise in being an effective screening tool to expedite the expansion of screening programs in low-resource settings.

\section{Acknowledgments}

We acknowledge the partial support of Consortium for Affordable Medical Technologies (CAMTech) at Massachusetts General Hospital with funds provided by the generous support of the American people through the United States Agency for International Development (USAID Grant number 224581) for collecting the images for this study. We would also like to acknowledge the support of Mark Schiffman, M.D, M.P.H., Division of Cancer Epidemiology and Genetics, National Cancer Institute, USA, for providing us with cervix images from the NIH database.  We also would like to acknowledge the support of IARC for sharing the  cervix image atlas dataset with us for this study.

\section{Declaration of interest}

The authors declare that they have no conflict of interest. This research did not receive any specific grant from funding agencies in the public, commercial, or not-for-profit sectors. 


\printcredits



\bibliographystyle{elsarticle-num}
\bibliography{paper}

\end{document}